\documentclass[preprint,showpacs,amsmath,amssymb]{revtex4} 
\usepackage{graphicx}
\usepackage{psfrag}

\begin{document}
\title{Bona fide Stochastic Resonance: A view point from  
 stochastic energetics.}
\author{Debasis Dan}
\email{dan@iopb.res.in}
\author{A. M. Jayannavar }
\affiliation{Institute of Physics, Sachivalaya Marg,
           Bhubaneswar 751005, India.}
\date{\today}
\begin{abstract}
        We investigate the resonance type behaviour of an overdamped Brownian
particle in a bistable potential driven by external periodic signal. It
has been shown previously that the
input energy pumped into the system by the external drive 
shows resonance type behaviour as function of
noise strength. We further extend this idea to study the behaviour as 
function of frequency of the external driving force and show the occurrence
of similar nonmonotonic behaviour, which can be ascribed as a signature
of bona fide stochastic resonance. Both weak and strong driving limit has
been explored indicating the occurrence of marginal supra-threshold stochastic
resonance in a bistable potential system. 
\end{abstract}
\maketitle

\section{Introduction}
  Stochastic Resonance (SR) is a nonlinear phenomena where the combined
effect of noise and nonlinearity (bistable systems or threshold systems)
leads to  an enhanced response of a weak periodic signal with addition 
of noise of optimal intensity. This counter intuitive phenomena has
been observed experimentally and in numerical and analog
simulations~\cite{review}.  
        However, there has been a lot of discussions in recent years 
regarding the validity
of SR  as bona fide stochastic resonance~\cite{contro}. The  
archetype of SR models is represented by a simple symmetric bistable
potential driven by a zero mean Gaussian white noise and an external 
sinusoidal bias. The response of such system has been mainly characterized 
by the response amplitude (RA) of the periodic component of the
process $< x(t) >$ or 
the signal to noise ratio (SNR). Both these response functions show typical
resonance type feature with increasing noise amplitude $D$~\cite{review}. 
This resonance
has been attributed to the matching of escape rate across the barrier
and the frequency of the external periodic drive. An obvious conclusion
is the occurrence of a similar peak with increasing frequency of the bias.
But such nonmonotonic behaviour has not been observed in spite of
exhaustive numerical and analytical studies on SNR and RA. However,
ref.~\cite{gitt} shows 
the existence of peak in SNR with increasing frequency for a system with 
rectangular potential barrier and for a special type of driving. 

        Other criteria have been proposed to study the resonance 
behaviour as function of frequency, thus characterizing SR as a bona
fide resonance. 
Gammaitoni et. al. have  shown that residence time distribution 
N(T) has a resonant behaviour 
as function of forcing frequency. Marchesoni et. al. recently showed 
numerically that in a Schmitt trigger N(T) shows a peak with frequency 
for both weakly and strongly driven system. The relationship between SR 
and synchronization of passages from one well to another can also be
characterized by hysteresis loop area~\cite{hyster,hyster2}. 
This loss can be taken as a measure of SR and SR is shown as a 
bona fide resonance. Recently it is argued that 
input energy is also a good measure of SR~\cite{iwai}. This energy is
equivalent to the work done by the external agent which drives the
potential periodically. The input energy  not only
shows peaking behaviour with temperature (noise strength) but also
takes into account only the interwell behaviour.  
In the conventional SNR both intrawell as well as interwell motion is taken 
into account and hence for small driving frequency and noise strength the 
motion is predominantly dominated by the intrawell oscillations~cite{iwai}. 
Hence the peak in the input energy 
is a better indicator of matching of escape rate and the external driving
frequency. This assertion has been made by taking into account the 
detailed comparison 
between various measures of SR. 

                In this work we  show that input energy not only
correctly shows the matching condition for noise induced escape rate 
and the external periodic drive, but resonance type behaviour is also 
obtained as function of $\omega$ (frequency of external drive), a 
signature of Bonafide SR. 
Bonafide SR is a relatively new term in stochastic dynamics, by which 
one means the actual matching of the time scales $T_{es}$ (barrier escape
time) and  
$T_{\omega}$ (period of external drive, $2\pi/\omega$). This is manifested 
as a peak in the plot of input energy  
with $T_{\omega} \mbox{ and } D$. The existence of 
SR in supra-threshold regime has also been demonstrated. We make an 
exhaustive study throughout the parameter regime of our problem and show the
detailed behaviour of SR peak as function of $D, \omega \mbox{ and } A$ in 
our model. 

\section{Model}
        We consider an overdamped Brownian particle moving in a bistable
potential $V(x,t) = -x^{2}/2 + x^{4}/4 -Ax\sin(\omega t)$ under the
influence of a zero mean white Gaussian noise $\xi(t)$ with
correlations $ <\xi(t)\xi(t')> = 2D\delta(t-t')$. The 
Langevin equation for such system is 
\begin{equation}
\label{langv}
\dot x(t) = -V'(x,t) + \xi(t) 
\end{equation}
and the corresponding Fokker Planck equation (FPE) is 
\begin{equation}
\label{fpe}      
\frac{\partial P(x,t)}{\partial t} = -\frac{\partial}{\partial x}
\big( V'(x,t) - D\frac{\partial}{\partial x})P(x,t),
\end{equation}
where $P(x,t)$ is the probability density of the particle at position
$x$ at time $t$.  
The barrier height $\Delta V = 0.25$. We consider both weak forcing
$Ax_{m} < \Delta V$ as well as strong forcing $Ax_{m} > \Delta V$
limit. Depending on the parameter regime, the Brownian particle is
either dominated by intra well or inter well oscillations. For small
driving force the motion is mostly dominated by intrawell
oscillations. At certain optimal noise strength the interwell
motion is enhanced due to the combined effect of noise and weak
periodic modulation. Such resonant enhancement of interwell motion is
termed as Stochastic Resonance (SR).  In this particular case the oscillatory
driving force keeps the system away from equilibrium. The energy
required to drive the system ($E_{in}$) can be calculated from
Sekimoto's stochastic energetics formalism~\cite{energy} and its has been shown that 
$E_{in}$ depends nonmonotonically on the noise strength $D$. Since the 
mean current in the system is zero, hence no work is done by the
system and all the input energy is dissipated into the bath. 
This suggests that input energy is related to dissipative loss or 
hysteresis loss in the system~\cite{hyster,hyster2}. 
Our effort is to find 
the signatures of SR in the input energy of the system as function of
system parameters $\omega \mbox{ and } D$. 

   The input energy per period($T_{\omega}$) is defined as ~\cite{energy,iwai}
\begin{equation}
\label{input}
E_{in} = \int_{t_{0}}^{t_{0}+T_{\omega}} < \frac{\partial V(x,t)}{\partial t}
dt> = -A\omega \int_{t_{0}}^{t_{0}+T_{\omega}} <x(t)\cos(\omega t) dt>. 
\end{equation}

The average $<..>$ is done over an ensemble of particles. The asymptotic 
probability distribution $P(x,t)$ is calculated by solving the FPE 
(\ref{fpe}). The input energy $E_{in}$ can be rewritten  in terms of asymptotic 
distribution as 

\begin{equation}
\label{main}
E_{in} = \int _{-\infty}^{\infty}dx \int_{t_{0}}^{t_{0}+T_{\omega}}dt~  
x(t)\cos(\omega t)P_{asy}(x,t). 
\end{equation}
where $P_{asy}(x,t)=P_{asy}(x,t+T_{\omega})$. This distribution is obtained 
after the initial transients have died down and asymptotically
probability density assumes a unique limiting periodic distribution in 
time. 
The maxima of $E_{in}$ as function of $D$ is taken as a signature of SR. 
We numerically solve eqn. (\ref{fpe}) by method of finite difference to obtain 
the asymptotic distribution $P_{asy}(x,t)$. $E_{in}$ is obtained by 
numerically integrating eqn. (\ref{main}). Throughout this 
work all the physical quantities are in dimensionless unit~\cite{review}.

\section{Results and Discussions}
        
\begin{figure}[h]
\includegraphics[width=8.6cm]{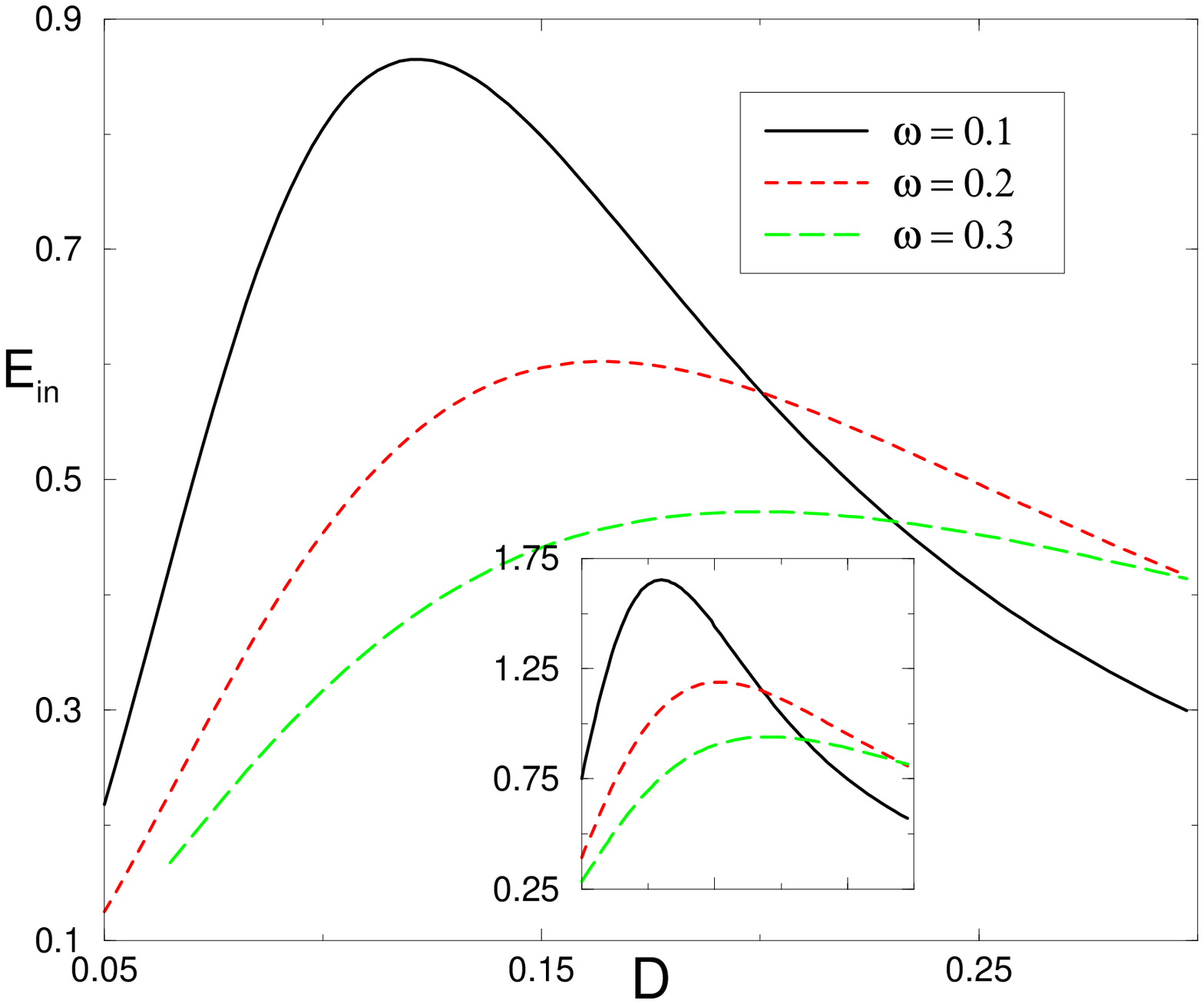}
\caption{\label{Ein-D}}
\end{figure}

\begin{figure}[h]
\includegraphics[width=8.6cm]{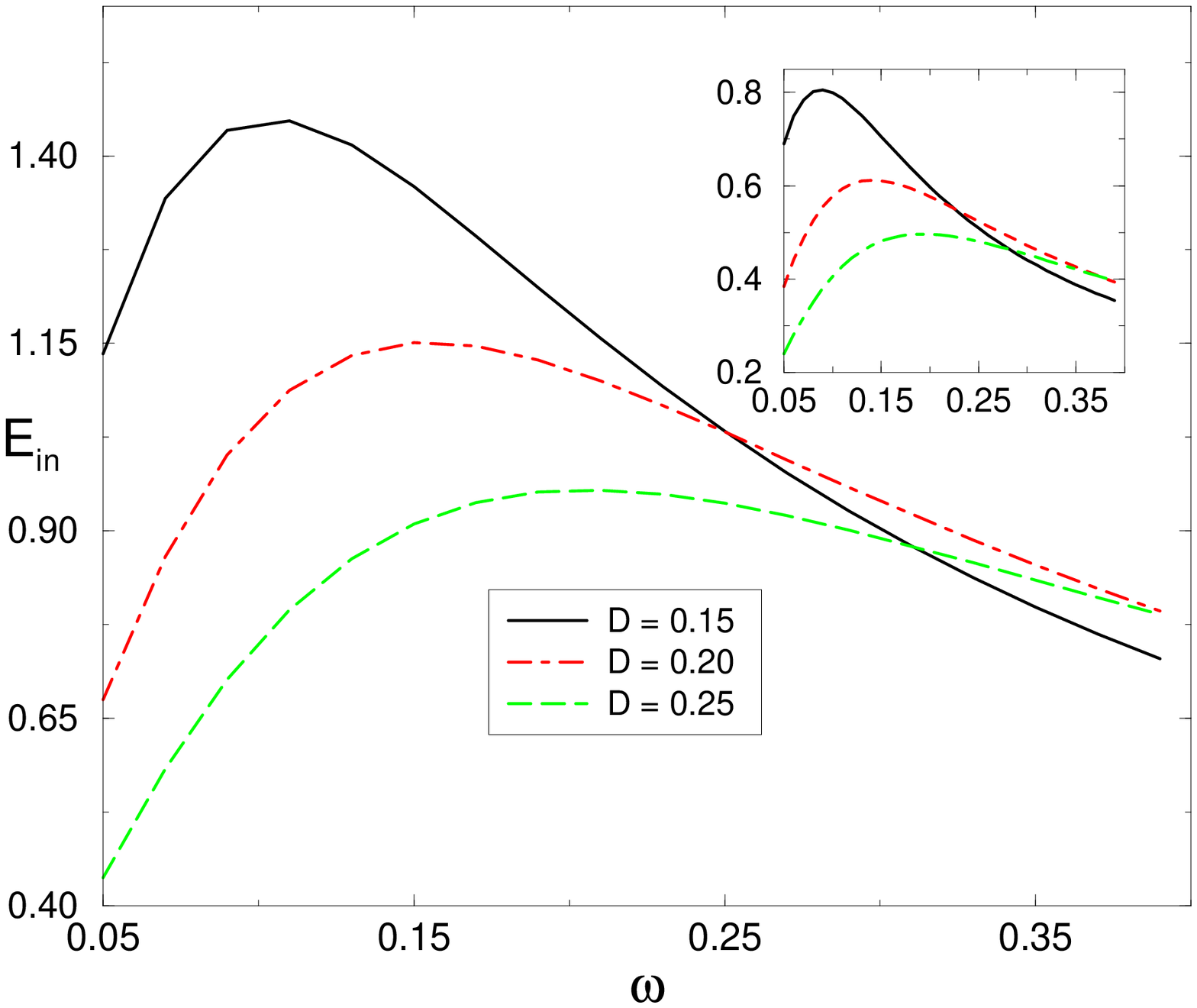}
\caption{\label{Ein-omg}}
\end{figure}

        The input energy shows a rich structure as function of $A, \omega 
\mbox{ and } D$. We mainly concentrate on low amplitude drive, $Ax_{m}/\Delta
 < 1$. In this regime input energy shows a sharp maxima both as function of
$D \mbox{ and } \omega$. In fig.~\ref{Ein-D}  we plot $E_{in}$ as function 
of $D$
for $A = 0.1$ and different values of $\omega$. The peak in the input energy 
can be attributed to the synchronization of escape from the potential well and
the external periodic drive as has been extensively discussed in previous 
literatures \cite{iwai}. With increasing $\omega$  the temperature at which 
$E_{in}$ peaks increases as shown in fig.~\ref{Ein-D}. The most favourable 
condition 
for hopping to other potential well is when the barrier height is minimum. 
When $\omega$ is increased the Brownian particle spends less time in the 
most favourable condition and hence stronger fluctuations are needed to cross
the barrier. Barrier height also decreases with increase in amplitude of 
external drive. Hence for reasons similar to above the resonance peak shift
towards lower temperature as shown in the inset, where we have plotted 
$E_{in}$ \textit{vs} $D$ for various values of $A$ as mentioned in the caption. 
However, our main motive is to check 
whether these resonance features are also observed when the frequency of the 
external drive is varied. In fig.~\ref{Ein-omg} we plot $E_{in}$ as function
of $\omega$ for 
$A=0.1$ and different values of temperature. Unlike other response function 
like SNR, RA which characterize SR and has monotonic dependence on $\omega$
, input energy shows a nonmonotonic and a peaking behaviour with $\omega$.  
For small amplitude drive and at low temperature the resonant frequency is very
close to half the Kramer's rate for the unperturbed system ( the resonance 
condition being $1/r_{k} = T_{\omega}/2 \rightarrow \omega/\pi = r_{k}$). 
For $D=0.2$, the Kramer's rate 
($r_{k} = \frac{1}{\sqrt 2 \pi}\exp(-0.25/D)$) is $0.0645$, which is 
very close to $\omega/\pi = 0.14/\pi = 0.0446$. 
With higher amplitude the resonance peak shifts to higher frequency for 
a given temperature as shown in the inset. Thus SR as a bona fide resonance 
is established. 
        
\begin{figure}[h]
\includegraphics[width=8.6cm]{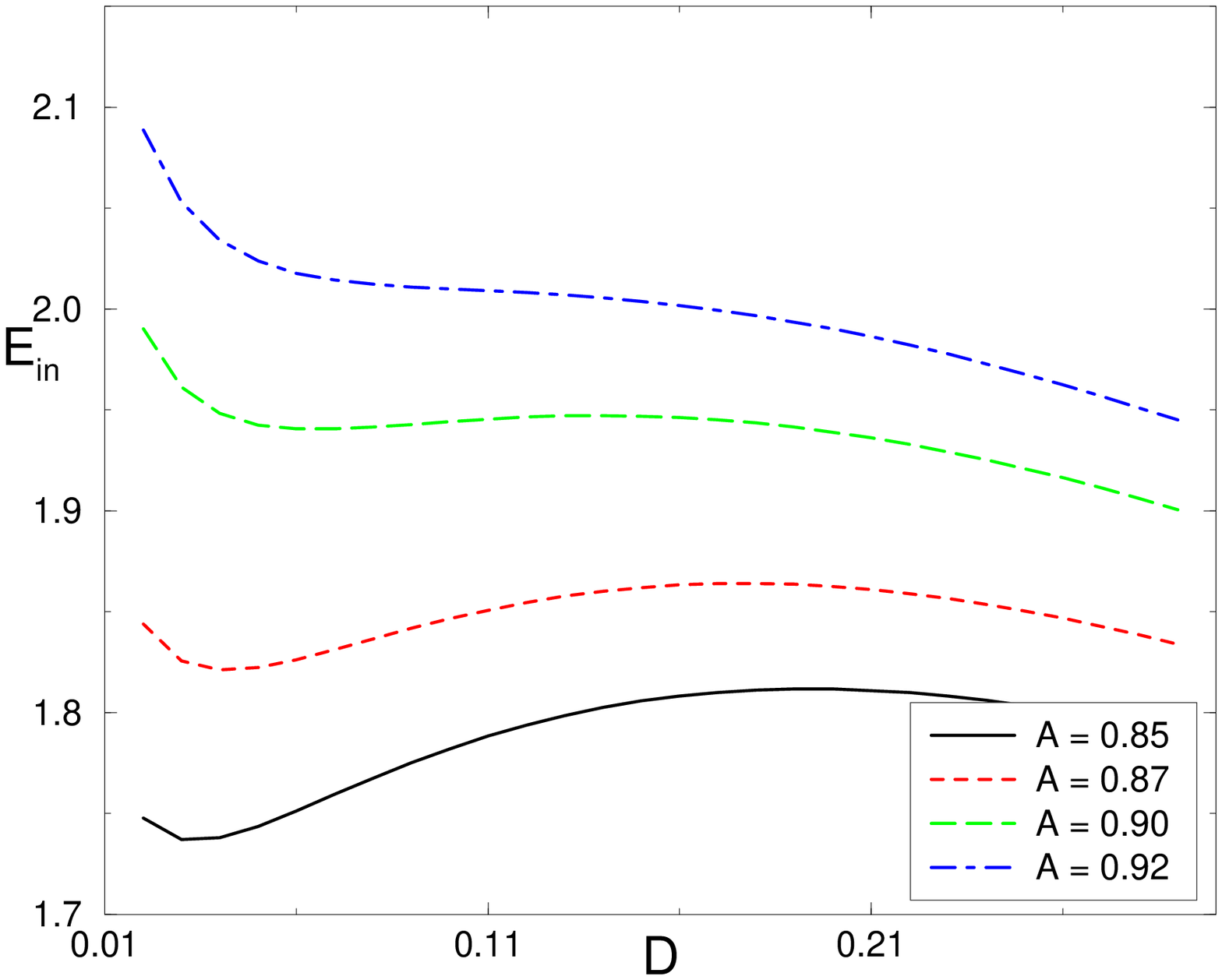}
\caption{\label{supra}}
\end{figure}

\begin{figure}[h]
\includegraphics[width=12.6cm]{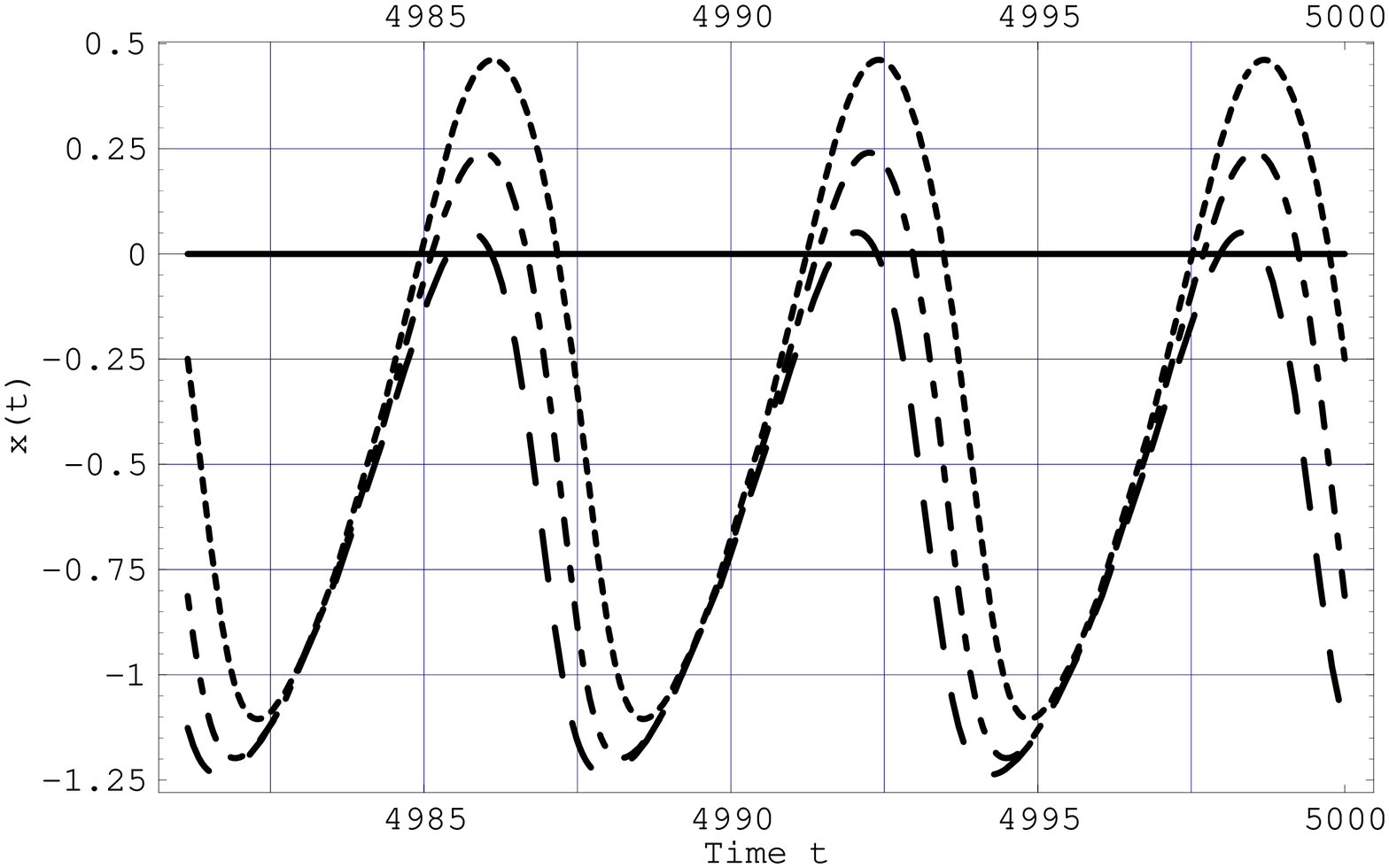}
\caption{\label{trajec}}
\end{figure}

        It is known that for $A >> A_{max}$, where $A_{max}$ is the 
dynamical threshold above which deterministic switch events take place
driven by the periodic signal alone (absence of noise), SR is not observed. 
The dynamical threshold depends 
on both the modulation frequency and the wave form $A(t)$. For the form 
given in eqn.(~\ref{langv}) $A_{max}/A_{th} = 1 + \beta(\Omega/a)$, where
$\beta = (2\sqrt3/\pi)g_{1}$ and $g_{1}$ is the smallest zero of the Airy
function $Ai'(-x)$, is order of unity \cite{switch}. We investigate
the marginal supra-threshold regime ($A > A_{max}$) and show that SR is observed
with increasing temperature provided $A$ is not very large compared to 
$A_{max}$. The occurrence of SR in this regime is related to noise
induced stability~\cite{hyster2,switch}. In fig.~\ref{supra} we plot 
input energy for four different valued of
$A$ at $\omega = 1.0$. Figure~\ref{trajec} shows the corresponding 
deterministic trajectories. As shown in fig.~\ref{trajec}, $A = 0.85$ is the 
threshold, i.e, the particle is just able to cross the barrier at $x=0$. 
Hence higher values of $A$ is above dynamic threshold. 
For $A > 0.85$ the input energy shows a resonance peak with increasing
temperature. This peak in $E_{in}$ in supra-threshold regime is shown for a very small 
window of $A$. For the specific parameter values as above, the peak vanishes 
for $A > 0.91$. This peak can be ascribed to resonant trapping as discussed 
in previous literature~\cite{hyster2,switch}. However, in  
supra-threshold regime $E_{in}$ clearly exhibits peaking behaviour as 
function of $\omega$~\cite{supra} for all values of $A$. This is more 
akin to conventional resonance (absence of barrier). 


        In conclusion we have calculated that the input energy pumped into 
the system by an external drive using the method of stochastic
energetics. This input energy is  
shown to be a good quantitative measure of SR. Moreover SR is shown to be a 
bona fide resonance. 

\newpage
\hspace{1.5in} \large  FIGURE CAPTIONS .\vspace{0.5in} \\
\large
Fig 1. $E_{in}$ \textit{vs} $D$ for $A = 0.1$. The inset 
shows the plot of $E_{in}$ \textit{vs} $D$ for $\omega = 0.1,~ 0.2,~ 0.3$ 
from top to bottom at $A = 0.2$. 

\vspace{.3cm}
Fig. 2. $E_{in}$ \textit{vs} $\omega$ for $A = 0.1$. Inset is the 
plot of $E_{in}$ \textit{vs} $\omega$ at $A = 0.2$ for $D = 0.15,~ 0.2,~
 0.25$ from top to bottom curve respectively. 

\vspace{.3cm}
Fig. 3. $E_{in}$ \textit{vs} $D$ at $\omega = 1.0$ for large amplitude 
$A = 0.85,~ 0.87,~0.90 \\ \mbox{ and } 0.92$ respectively. 

\vspace{.3cm}
Fig. 4. The determistic trajectories ($D = 0$) for amplitudes \\ $A = 0.85,~
0.90 \mbox{ and }0.92$. All these trajectories cross the barrier at $x = 0$.

\newpage 

\end{document}